\documentclass[compsoc, conference, a4paper, 10pt, times]{IEEEtran}

\usepackage[utf8]{inputenc} 
\usepackage[T1]{fontenc}    
\usepackage[hyphens]{url}
\usepackage{cite}
\usepackage{amsmath,amssymb,amsfonts}
\usepackage{algorithmic}
\usepackage{graphicx}
\usepackage{textcomp}
\usepackage{xcolor}
\usepackage{booktabs}
\usepackage[hidelinks]{hyperref}
\usepackage{subcaption}
\usepackage{amsmath}
\usepackage{amsfonts}
\usepackage{tikz}
\usepackage{multirow}
\usepackage{algorithm}
\usepackage{optidef}
\usepackage{comment}

\begin{document}

\title{Query-Free Evasion Attacks Against Machine Learning-Based Malware Detectors with Generative Adversarial Networks}


\iftrue
\author{\IEEEauthorblockN{Daniel Gibert}
\IEEEauthorblockA{\textit{CeADAR, University College Dublin} \\
Dublin, Ireland\\
daniel.gibert@ucd.ie}
\and
\IEEEauthorblockN{Jordi Planes}
\IEEEauthorblockA{\textit{University of Lleida} \\
Lleida, Spain \\
jordi.planes@udl.cat}
\and
\IEEEauthorblockN{Quan Le}
\IEEEauthorblockA{\textit{CeADAR, University College Dublin} \\
Dublin, Ireland\\
quan.le@ucd.ie}
\and
\IEEEauthorblockN{Giulio Zizzo}
\IEEEauthorblockA{\textit{IBM Research Ireland} \\
Dublin, Ireland\\
giulio.zizzo2@ibm.com}
}
\fi

\maketitle

\begin{abstract}
Malware detectors based on machine learning (ML) have been shown to be susceptible to adversarial malware examples. However, current methods to generate adversarial malware examples still have their limits. They either rely on detailed model information (gradient-based attacks), or on detailed outputs of the model - such as class probabilities (score-based attacks), neither of which are available in real-world scenarios. Alternatively, adversarial examples might be crafted using only the label assigned by the detector (label-based attack) to train a substitute network or an agent using reinforcement learning. Nonetheless, label-based attacks might require querying a black-box system from a small number to thousands of times, depending on the approach, which might not be feasible against malware detectors.

This work presents a novel query-free approach to craft adversarial malware examples to evade ML-based malware detectors. To this end, we have devised a GAN-based framework to generate adversarial malware examples that look similar to benign executables in the feature space. To demonstrate the suitability of our approach we have applied the GAN-based attack to three common types of features usually employed by static ML-based malware detectors: (1) Byte histogram features, (2) API-based features, and (3) String-based features. Results show that our model-agnostic approach performs on par with MalGAN, while generating more realistic adversarial malware examples without requiring any query to the malware detectors. Furthermore, we have tested the generated adversarial examples against state-of-the-art multimodal and deep learning malware detectors, showing a decrease in detection performance, as well as a decrease in the average number of detections by the anti-malware engines in VirusTotal. 
\end{abstract}



\section{Introduction}
\label{sec:introduction}
The rate of cybercrimes is increasing every year, and their cost to the world is estimated to be \$8 trillion annually in 2023, representing the greatest transfer of economic wealth in history.\footnote{\url{https://cybersecurityventures.com/cybercrime-to-cost-the-world-8-trillion-annually-in-2023/}} There are many types of cyberattacks, including Denial-of-Service (DoS) attacks, phishing, malicious software or malware, SQL injection, and zero-day exploits. Among the aforementioned attacks, malware, and more specifically ransomware, has reached epidemic proportions globally, with an estimated cost of \$20 billion in 2021.\footnote{\url{https://cybersecurityventures.com/global-ransomware-damage-costs-predicted-to-reach-20-billion-usd-by-2021/}}

To defend against malware, a layered defense is typically employed, with various layered security elements working in conjunction with each other to keep computer devices safe. 

One of the most important components is endpoint protection. Traditionally, endpoint protection has relied on signature-based antivirus solutions to detect malware, consisting of a large database of malicious software signatures and definitions. These solutions detect malware by scanning files and looking for patterns that match the signatures and definitions from the database. As a result, they can only recognize known threats. To mitigate new, unknown threats, endpoint protection solutions started adopting machine learning as it has proven capable of discovering hidden patterns from huge amounts of data without human intervention\cite{GIBERT-JNCA-20201}. Nowadays, most modern anti-virus solutions, also known as Next-Generation Antivirus (NGAV), use a combination of machine learning and behavioral detection so that known and unknown threats can be mitigated and immediately prevented.

\subsection{Motivation}
Unfortunately, machine learning-based malware detectors can be fooled by evasion attacks, where the goal of the attacker is to modify a given executable in order to evade detection. These carefully crafted executables that evade detection are referred to as adversarial malware examples. Various approaches~\cite{DBLP:conf/eusipco/KolosnjajiDBMGE18,DBLP:journals/corr/abs-1802-04528,DBLP:conf/sp/Al-DujailiHHO18,9437194,YUSTE2022102643,DBLP:journals/corr/abs-1801-08917,DBLP:journals/corr/HuT17} to generate adversarial malware examples have been presented in the literature. The majority of these attacks rely either on complete knowledge of the model~\cite{DBLP:conf/eusipco/KolosnjajiDBMGE18,DBLP:journals/corr/abs-1802-04528,DBLP:conf/sp/Al-DujailiHHO18}, i.e. gradient-based attacks, or on confidence scores such as the probability of the executable being malicious~\cite{9437194,YUSTE2022102643}, i.e. score-based attacks. However, in a real-world scenario only the decision of the detector is available~\cite{DBLP:journals/corr/abs-1801-08917,DBLP:journals/corr/HuT17}, i.e whether the executable is malicious. One approach to attack a black-box detector in this setting is to use the labels to train a substitute detector that emulates the black-box detector and then attack the substitute detector~\cite{DBLP:journals/corr/HuT17}. Another approach in this setting is to train an agent using reinforcement learning to select which set of actions to perform on a Portable Executable (PE) file in order to evade detection~\cite{DBLP:journals/corr/abs-1801-08917}. Nonetheless, the aforementioned methods require from few to unlimited number of queries to attack the black-box detectors, which might raise suspicion concerning the submitted samples, as submitting a high number of similar queries, or any query, to a cloud security provider might result in a close and thorough inspection of the files. To make things worse, the aforementioned attacks assume detailed knowledge of the model's input features; which, given the secrecy and confidentiality of cybersecurity actors, is not available.

\subsection{Contributions}
Given the aforementioned limitations of evasion attacks against malware detectors, this paper presents a query-free end-to-end evasion attack that generates adversarial malware executables by exploiting the distinct characteristics of benign and malicious executables. The main contributions of this paper are the following:
\begin{itemize}
    \item We propose a general framework using Generative Adversarial Networks to generate adversarial malware executables. Our Conditional Wasserstein GAN generates malware examples that resemble benign examples in the feature space, thus fooling malware detection systems. The GAN architecture consists of two networks, the generator and the critic, that allow us to automatically generate fake malicious examples that look similar to real benign examples. 
    \item The generalization ability of our approach has been tested on three different type of features commonly employed by ML-based malware detectors to discern between goodware and malware: (1) the executables' byte distribution, (2) the libraries and functions imported, and (3) the strings found in the executables' content. For instance, the GAN-based framework will transform the malware's byte distribution into a more "benign" byte distribution according to the generator's output.
    \item We show how the attack performed in the feature space can be converted to an end-to-end attack. For example, to modify the executables' byte distribution one can append the corresponding bytes necessary to move from the original to the target byte distribution at the end of the Portable Executable files. This process is known as overlay append.
    \item We formulate the problem of determining the number of byte values to be appended at the end of executables as an integer linear programming problem with soft constraints on the byte frequency so that the size of the resulting adversarial malware executables is minimized to avoid unmanageable growth in size.
    \item We demonstrate on a public benchmark, the BODMAS dataset~\cite{bodmas}, that the proposed model agnostic attack performs on par with the MalGAN black-box evasion attack~\cite{DBLP:journals/corr/HuT17} without requiring any queries to the target malware detectors in order to craft the adversarial malware executables.
    \item We further analyze the evasion performance on state-of-the-art malware detectors, showing the transferability of our attacks.
    \item We upload the generated adversarial malware executables to VirusTotal to demonstrate the suitability of our attack to evade some commercial anti-virus solutions. 
\end{itemize}

\section{Related Work}
\label{sec:related_work}
Machine learning-based malware detectors have proven to be vulnerable to evasion attacks, adversarial attacks that consists of carefully perturbing the malicious executables at test time to have them misclassified as benign software. Evasion attacks in the literature can be categorized broadly in two groups, depending on the attacker's access to the model: (1) white-box attacks~\cite{DBLP:conf/eusipco/KolosnjajiDBMGE18,DBLP:journals/corr/abs-1802-04528,DBLP:conf/sp/Al-DujailiHHO18}, where an attacker has access to detailed information of the model such as the learning algorithm and its parameters, and (2) black-box attacks, where the attacker only has access to the output assigned to a given executable. Moreover, black-box attacks can be further divided into score-based~\cite{9437194,YUSTE2022102643} and label-based~\cite{DBLP:journals/corr/abs-1801-08917,DBLP:journals/corr/HuT17} attacks depending on whether the output of the model is a confidence classification score or a label indicating whether the executable is malicious or benign.

B. Kolosnjaji et al.~\cite{DBLP:conf/eusipco/KolosnjajiDBMGE18} introduced a gradient-based attack to generate adversarial malware executables by manipulating certain bytes in each executable to maximally increase the probability of the executables being classified as benign. The attack aims at minimizing the confidence associated with the malicious class under the constraint that $q_{max}$ bytes can be injected using gradient descent. The attack was conceived against MalConv~\cite{DBLP:conf/aaai/RaffBSBCN18}, a shallow convolutional neural network trained on raw bytes.

Kreuk et al.~\cite{DBLP:journals/corr/abs-1802-04528} and O. Suciu et al.~\cite{8844597} adapted the Fast Gradient Sign Method originally described in Biggio et al.~\cite{DBLP:conf/pkdd/BiggioCMNSLGR13} to generate adversarial malware executables. This was done by generating a small-sized adversarial payload and iteratively updating its bytes until the adversarial executables evade being detected by MalConv. Similarly, Al-Dujaili et al.~\cite{DBLP:conf/sp/Al-DujailiHHO18}, adapted a well-known gradient-based inner maximization methods for continuous feature spaces, the Fast Gradient Sign Method (FGSM), to binary feature spaces. As a result, the adapted version of FGSM could be used to modify a binary indicator feature vector. Each index of the feature vector represents a unique API function, where a "1" indicates the presence of the corresponding API function in the Import Address Table (IAT) of the PE executable.

L. Demetrio et al.~\cite{9437194} proposed a functionality-preserving black-box attack. It injects benign content, extracted from benign software, at the end of a malicious file or within newly-created sections. Afterwards, a genetic algorithm is used to modify the injected bytes until the resulting malicious file evades detection. Similarly, J. Yuste et al.~\cite{YUSTE2022102643} presented a score-based black-box attack based on dynamically introducing unused blocks, or section caves, within malware binaries. Afterwards, the content of the newly-introduced blocks of bytes is optimized using a genetic algorithm.

H. Anderson et al.~\cite{DBLP:journals/corr/abs-1801-08917} proposed a general framework for attacking static ML-based malware detectors via Reinforcement Learning (RL). In their work, they trained a RL agent using the Deep Q-Network algorithm to select the actions to perform on a PE file among a set of functionality-preserving operations including, but not limited to, adding a new function to the Import Address Table, manipulating section names, creating new sections, modifying the slack space between sections, packing or unpacking the file.

W. Hu et al.~\cite{DBLP:journals/corr/HuT17} introduced a GAN-based algorithm named MalGAN to generate adversarial API-based feature vectors to attack simple unimodal static API-based malware detection models. MalGAN consists of two feed-forward neural networks, (1) a generator and (2) a substitute detector. The generator network is trained to minimize the generated adversarial malware feature vectors' maliciousness probabilities predicted by the substitute detector. The substitute detector is trained to fit the API-based malware detection system. By training both networks together, the generator will learn what changes have to be performed to the malware's feature vector in order to evade the target API-based malware detection system.

\subsection{Limitations of Existing Adversarial Evasion Attacks}

Despite the aforementioned research, current methods used to generate adversarial malware examples are limited and not practical in the real-world. On the one hand, white-box or gradient-based attacks, although they successfully generate adversarial malware examples, are not feasible in a real-world scenario as the algorithm and parameters of the machine learning malware detectors are not available to attackers. In addition, the maliciousness score predicted by the ML-based detectors is not available to attackers either, and thus, score-based attacks are also not realistic. On the other hand, the only information that is provided by malware detectors is the label associated to a given executable, that is, whether or not the executable is malicious. However, current methods require from a small number of queries~\cite{DBLP:journals/corr/abs-1801-08917} to unlimited queries to attack the black-box detectors~\cite{DBLP:journals/corr/HuT17}, which might raise suspicion on the submitted samples. For instance, VirusTotal~\footnote{\url{https://www.virustotal.com}}, a popular aggregator scanner, shares the submitted files between the examining partners, who use the results to improve their own systems. Furthermore, submitting similar queries, or any query, to a cloud security provider might generate suspicion and raise the alarm, resulting in a close and thorough inspection. Given the aforementioned constraints, we propose a \emph{query-free} evasion attack to craft adversarial malware examples based on the assumption that the more similar the malicious executables are to benign executables in terms of structure and behavior, the harder it is for the ML-based detector to classify them correctly.

\section{Towards Generating Adversarial Malware Examples with GANs}
\label{sec:methodology}
This paper proposes a query-free approach to generating adversarial malware examples without assuming any known knowledge of the target malware detection system we want to evade, including the input features, the machine learning algorithm used, the parameters of the model, or the output of the model. Instead, this work is based on the assumption that malicious and benign pieces of software are inherently different, in terms of structure and behavior, which machine learning algorithms such as boosted decision trees or neural networks take advantage of to learn a function mapping the input variables to a target label. For malware detection, the input variables are the features extracted either statically or dynamically from the executables and the output is the probability that an executable is malicious. Thus, by altering the malicious executables in a way that they resemble benign software we might evade detection.

\begin{figure}[ht]
	\centering
	\usetikzlibrary{shapes.geometric, arrows.meta, backgrounds, fit, positioning}
	\usetikzlibrary{matrix,chains,scopes,positioning,arrows,fit}
	\tikzstyle{software} = [rectangle, rounded corners, minimum width=1.5cm, minimum height=1cm,text centered, draw=black, fill=white]
	\tikzstyle{features} = [rectangle, minimum width=1cm, minimum height=0.5cm,text centered, draw=black, fill=white]
	\tikzstyle{classifier} = [diamond, text centered, draw=black, fill=white, text width=1.7cm] 
	\tikzstyle{y_pred} = [rectangle, text centered, draw=black]

	\tikzstyle{arrow} = [thick,->,>=stealth]
	\tikzstyle{arrow1} = [thick,<-=stealth]
	\resizebox{0.9\columnwidth}{!}{
		\begin{tikzpicture}
		\node (goodware) [software, xshift=4cm] {Goodware};
		\node (malware) [software, yshift=-2cm] {Malware};
		\node (benign_features) [features, rotate=90, right of=goodware, yshift=-2cm, xshift=-1cm] {Benign features};
		\node (malicious_features) [features, rotate=90, right of=malware, yshift=-2cm, xshift=-1cm] {Malicious features};
		\node (noise) [features, rotate=90, left of=malicious_features, xshift=-1cm] {Noise};
		\node (generator) [classifier, right of=malicious_features, xshift=1cm, yshift=-0.5cm] {Generator};
		\node (fake_features) [features, rotate=90, below of=generator, yshift=-1cm] {Fake features};
		\node (critic) [classifier, right of=benign_features, xshift=1cm, yshift=-1cm] {Critic};

		\node (downdisc) [draw=none, below of=critic, yshift=-2.5cm]{};
		\node (downgen) [draw=none, below of=generator, yshift=-1cm]{};

		\draw[->] (goodware) -- (benign_features);
		\draw[->] (malware) -- (malicious_features);
		\draw[->] (malicious_features) -- (generator);
		\draw[->] (noise) -- (generator);
		\draw[->] (generator) -- (fake_features);

		\draw[->] (benign_features) -- (critic);
		\draw[->] (fake_features) -- (critic);
		
        \draw[->] (critic) -- (downdisc.south) -- (downgen.south) -- (generator);
		\end{tikzpicture}
	}
	\caption{The architecture of the GAN.}
	\label{fig:gan_architecture}
\end{figure}
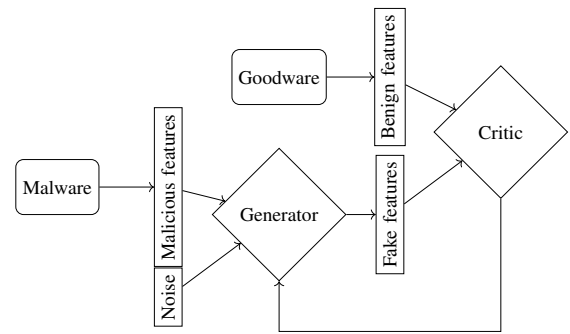

Recently, a class of machine learning algorithms, named Generative Adversarial Networks or GANs~\cite{DBLP:conf/nips/GoodfellowPMXWOCB14}, has been proposed to generate fake data that are similar to real data. GANs are an approach to generative modelling using deep learning methods. A GAN consists of two neural networks, namely the generator and the discriminator, which are in competition with each other in order to discover the patterns or regularities in the given real data in such a way that the generator learns to generate fake data or new examples that plausibly could have been drawn from the original dataset. In our work, we use a Conditional Wasserstein GAN~\cite{DBLP:conf/icml/ArjovskyCB17} to generate adversarial malware examples similar to benign executables. To this end, the GAN will be used to transform a malicious feature vector in a way that it resembles a benign feature vector without altering the original malware's behavior. See Figure~\ref{fig:gan_architecture} for a complete description of the GAN architecture.

\subsection{Feature Types}
In this work, we apply GANs to three types of features commonly used by ML-based malware detectors:
\begin{itemize}
    \item The byte frequency distribution of the executables, referred to as \emph{byte unigram features}.
    \item The libraries and functions imported by the executables, referred to as \emph{API-based features}.
    \item The ASCII strings found in the executables' content, referred to as \emph{string-based features}.
\end{itemize}
\subsubsection{Byte Unigram Features}
The simplest and most common type of features usually extracted from executables to detect malware is the byte frequency distribution of the executables, also known as byte unigram features. Byte unigram features represent the frequency of each byte in the executable, and thus, are described with a 256-dimensional vector. Mathematically, the byte unigram features of a given executable $x$ can be described as follows:

$$ \mathbf{x} = \begin{bmatrix} 
           x_{0} \\
           x_{1} \\
           \vdots \\
           x_{255}
         \end{bmatrix}\: 
\text{where} \sum_{i=0}^{255} x_{i} = 1.0
$$


\subsubsection{API-based Features}
The Application Programming Interface (API) provide services to other pieces of software to communicate with each other and to communicate with the hardware of a computer system. Although the use of the operating system (OS) API is not illegitimate by itself, malware writers make use of these API functions to interact with the OS and perform nefarious tasks.
The libraries and functions imported by executables are usually mapped as a binary feature vector, $x \in \{0,1\}^M$. More specifically, if $M$ API functions are used as features, an $M$-dimensional feature vector is constructed to represent a given executable. If the executable imports the $d$-th API function, the $d$-th feature value is set to 1; otherwise it is set to 0.

\subsubsection{String-based Features}
Strings are ASCII and Unicode-printable sequences of characters embedding within a file. Strings can give us information about the program functionality and indicators associated with malicious or suspicious behavior. For instance, strings extracted from a binary executable might contain references to filenames, URLs, domain names, IP addresses, registry keys, attack commands, etcetera.
The strings extracted from binary executables are also typically mapped as a binary feature vector, $x \in \{0,1\}^N$, where $N$ is the number strings used as features.

\subsection{Generator Network}
The generator receives as input the concatenation $c$ of a feature vector $m$ and a noise vector $z$. The size of $m$ and $z$, as well as the network architecture, depends on the type of features that we want the generator to generate. The idea behind feeding the original features to the generator is to condition it to craft a specialized adversarial feature vector~\cite{DBLP:journals/corr/MirzaO14}. $z$ is a $Z$-dimensional vector ($Z$ is different for each feature type), where each element of $z$ is a random number sampled from a uniform distribution in the range $[0,1)$. $z$ allows the generator to produce a wide variety of adversarial examples from a single malicious feature vector by sampling from different places in the input distribution. The input vector $c$ is fed into the generator, a multi-layer feed-forward neural network, which will generate an output vector denoted by $m'$. Depending on the input features the architecture of the generator might vary. See Table~\ref{tab:unigram_generator_configurator_details} for a complete description of the architecture. Below, we describe the main differences between the generator networks devised for each feature type.

\subsubsection{Byte Unigram Generator Network}
To generate a target byte frequency distribution that resembles those found in benign executables, the generator will receive as input a 256-dimensional vector $m$, where each element of $m$ corresponds to the frequency of a particular byte in the executable, concatenated with the noise vector $z$. To sum up, $m_{0}$ corresponds to the frequency of the byte 0x00, $m_{1}$ corresponds to the frequency of the byte 0x01, and so on. The output of the generator, whose architecture is specified in Table~\ref{tab:unigram_generator_configurator_details}, is also a 256-dimensional vector (same size as the input feature vector). That is, the output layer of the generator has 256 neurons and the activation function used by the last layer is the softmax function to force the generated features to sum to 1.

\subsubsection{API-based and String-based Generator Networks}
The API-based and String-based generator networks main difference with the byte unigram-based generator is the output layer. The input binary feature vector $m$ will have size $M$, which is the number of API functions or strings, respectively, used as input. The output layer of the generator network, denoted by $o$, will use the sigmoid function instead of the softmax function. Furthermore, a binarization transformation will be applied to $o$ according to whether or not the element is greater than $0.5$, which produces a binary vector $o^{'}$. In this case, we cannot freely modify all binary features as removing a feature from the original executable might break it. For this reason, we only allow new features to be added. The resulting adversarial feature vector can be expressed as $m^{'} = m | o^{'}$, where $|$ is the element-wise binary OR operation.

To back propagate the gradients we used the smooth function $G$ shown in Equation 1 that was defined by W. Hu et al.~\cite{DBLP:journals/corr/HuT17}. The smooth function was defined as follows:
\begin{equation}
G_{\theta}(m,z) = max(m, o)
\end{equation}
The idea behind $G$ is to use the network's real output value if an element of $m$ has value 0. Otherwise, it is 1. For more information about the smooth function $G$ we refer readers to the work of W. Hu et al.~\cite{DBLP:journals/corr/HuT17}.

\subsection{Critic Network}
The critic network receives as input a feature vector $x$, where $x$'s size depend on the feature type and outputs a ``benignness" score for a given sample. Vanilla GANs use a sigmoid activation function in the output layer of the discriminator to predict the likelihood of a given sample being real. Instead, the critic network replaces the sigmoid function with a linear function to predict the ``realness" for a given sample. In our case, this is the ``benignness". The critic network is a multi-layer feed-forward neural network. Cf. Table~\ref{tab:unigram_critic_configurator_details} for the details of the critic architecture.

\subsection{Training the GAN}
Training the Conditional Wasserstein GAN with Gradient Penalty~\cite{10.5555/3295222.3295327} to generate "benign" feature vectors requires collecting both benign and malicious executables, whereas the more representative the live malware and the benign software the better. 

The loss function of the critic $L_D$ is defined as:
\begin{equation}\label{eq:L_D}
\begin{split}
    L_{D} = \mathbb{E}_{\tilde{x}\sim\mathbb{P}_{g}}[f(\tilde{x})] - \mathbb{E}_{r\sim\mathbb{P}_{r}}[f(x)]\\
    + \lambda \mathbb{E}_{\check{x}\sim\mathbb{P}_{\check{x}}}[(|| \bigtriangledown_{\check{x}}  f(\check{x}) ||_{2} -1)^{2}]
    \nonumber
\end{split}
\end{equation}
where the terms to the left of the sum are the original critic loss and the terms to the right of the sum are the gradient penalty. $\mathbb{P}_{\check{x}}$ is the distribution obtained by uniformly sampling along a straight line between the benign and the generated distributions, $\mathbb{P}_{r}$ and $\mathbb{P}_{g}$, respectively. $\lambda$ is the penalty coefficient used to weight the gradient penalty term. In our experiments, we set $\lambda=10$.

To train the critic network, $L_{D}$ should be minimized with respect to the weights of the critic network. Instead of predicting the probability of a generated sample being ``benign", the critic in a Wasserstein GAN scores the ``benignness" or ``maliciousness" of a given feature vector. Unlike the vanilla GAN discriminator model that, once trained, may fail to provide useful gradient information for updating the generator model, the critic's loss does not saturate and hence always yields useful gradient information.

The loss of the generator is defined as:
\begin{equation}\label{eq:L_G}
    L_{G} = \mathbb{E}_{\tilde{x}\sim\mathbb{P}_{g}}[f(\tilde{x})]
    \nonumber
\end{equation}
where $\mathbb{P}_g$ is the generated distribution.

The whole process of training the GAN is shown in Algorithm~\ref{alg:gan_training}. For a given step in the training process, $\Theta_D$ is updated according to $L_D$, and for every $n\_generator$ step, so is $\Theta_G$ according to $L_G$.

\begin{algorithm}
\caption{Conditional Wasserstein GAN Training Process}
\begin{algorithmic} 
\REQUIRE B : set of goodware samples, M : set of malware samples, $\Theta_{D}$ : r weights, $\Theta_{G}$ : generator weights
\STATE $n\_generator \gets 5$
\STATE $MAX\_STEPS \gets |B| \times NUM\_EPOCHS$
\FOR{step $\gets$ 1 to  MAX\_STEPS}
    \IF{converged enough}
       \STATE break
     \ENDIF
    \STATE $b \gets sample\_minibach(B)$
    \STATE $m \gets sample\_minibatch(M)$
    \STATE $z \gets noise\_vector()$
    \STATE $m' \gets generator(m, z)$
    \STATE $\Theta_{D} \gets \Theta_{D} + \bigtriangledown_{\theta_{D}}L_{D}$
    \IF{$step \mod n\_generator = 0$}
        \STATE $m' \gets generator(m, z)$
        \STATE $\Theta_{G} \gets  \Theta_{G} + \bigtriangledown_{\theta_{G}}L_{G}$
    \ENDIF
\ENDFOR
\end{algorithmic}
\label{alg:gan_training}

\end{algorithm}

\section{From Feature-based to End-to-End}
So far, we have described the process followed to generate adversarial feature vectors with GANs. However, modifying the malware's feature vector representation is not an end-to-end attack. To convert the aforementioned feature-based attack into an end-to-end attack we need to modify the executables so that they have the generated adversarial features. Accordingly, the modifications that need to be performed to the executables depend on the type of features that we want to modify: (1) to modify the executables so as they have the target byte frequency distribution we will append the corresponding bytes at the end of the executables; (2) to add new libraries and import functions we will modify the Import Address Table of the executables; (3) to insert new strings we will create a new section and add the corresponding strings to it.
All the aforementioned modifications have been performed using LIEF\footnote{\url{https://lief-project.github.io/}}, a Python library specifically designed to parse and modify executables~\cite{2018arXiv180404637A,9437194,DBLP:journals/corr/abs-1801-08917}.

\subsection{Determining the Bytes to be Appended at the End of Executables}
\label{sec:determine_byte_values}
The problem of determining the number of byte values to be appended at the end of the executable to have a target byte frequency distribution in accordance with the original one can be codified as an integer linear programming problem:

\begin{mini!}|l|[3]
{}{\sum_{i=0}^{255}{p_{i}}}{}{} \nonumber
\addConstraint{b_i + p_i}{=
      r_i \times \left( \sum_{j=0}^{255}  b_j + p_j \right)}{\quad i = 0,\ldots,255}\nonumber
\end{mini!}
where $p_i$ is an integer variable that indicates the the amount of byte $i$ bytes that need to be padded at the end of the executable, $r_i$ is a real variable with the target byte distribution (ratio) we want to achieve for byte $i$, and $b_i$ is the original number of bytes found in the executable for byte $i$. However, this model results in huge padding values, due to the equality in the constraint, if a solution is found. The computation can be accelerated by relaxing the integer variables to real variables, which gives a good approximate solution (in the current experimentation, the  difference between the solutions is $\approx1$ byte).

However, appending bytes at the end of the executables to exactly map a target byte distribution from their original byte distribution generates large, unrealistic executables. To this end, we propose to map the original byte distribution to an approximated version of the target byte distribution, allowing for some error among the resulting byte unigram values. This can be done by allowing the solution to be near the required distribution in order to obtain lower (and practical) values. 
To map the original byte distribution to an approximated version of the target distribution, the constraint is changed by adding an upper bound and a lower bound both with a gap, the allowed error interval, as follows:

\begin{mini!}|l|[3]
{}{\sum_{i=0}^{255}{p_{i}}}{}{} \nonumber
\addConstraint{r_i \times \left( \sum_{j=0}^{255}  b_j + p_j \right)  - g \le b_i + p_i }{\quad i = 0,\ldots,255} \nonumber
\addConstraint{b_i + p_i \le r_i \times \left( \sum_{j=0}^{255}  b_j + p_j \right) + g}{\quad i = 0,\ldots,255} \nonumber
\end{mini!}
where $p_i, r_i$ and $b_i$ are as defined before, and $g$ is the gap, which is set to $0.001$ in the current experimentation.

The models have been implemented in \textsc{Zimpl},\footnote{\url{https://zimpl.zib.de/}} and solved with \textsc{SoPlex},\footnote{\url{https://soplex.zib.de/}} an optimization package for solving linear programming problems.
The authors are aware methods exist to detect whether or not the overlay of executables has been modified. However, as the goal of our work is to evade ML-based malware detectors we did not consider stealthier mechanisms to insert the new content, i.e. the byte values that must be inserted into the executable in order to have a specific target byte frequency distribution.

\section{Experiments}
\label{sec:experiments}

\subsection{Experimental Setup}
The dataset used in this paper is the BODMAS dataset~\cite{bodmas}. It consists of 57,293 malware and 77,142 benign Windows PE files. The dataset has been divided into training, validation, and testing sets, 
 consisting of 80\%, 10\% and 10\% of the data, respectively. The same training, validation and testing splits have been used to train our query-free GAN to generate adversarial examples by modifying the byte distribution, the Import Address Table and the Strings of malicious executables.

The experiments were run on a machine with an Intel
Core i7-7700k CPU, 1xGeforce GTX1080Ti GPU and 64Gb
RAM. The code has been implemented with PyTorch~\cite{NEURIPS2019_9015} and is publicly available in our Github repository~\footnote{\url{https://github.com/code_repository}.
It will be made available after the paper is accepted.}

\subsection{Attack Evaluation}
\label{sec:attack_evaluation}
This section presents the evaluation of the proposed query-free attack against various unimodal detectors, state-of-the-art detectors in the literature, and on VirusTotal Service.

\subsubsection{Attack Evaluation against Unimodal Detectors}
Our query-free attack has been evaluated against various unimodal malware detectors, i.e. they take as input a single type of features, to show the effect of our attacks in various scenarios. For each type of features we have trained one detector, using the samples from the EMBER dataset~\cite{2018arXiv180404637A}. Using only a single type of features will allow us to measure the evasion capability of our GAN-based approach. Below are listed the unimodal malware detectors evaluated against our GAN-based generated adversarial examples:
\begin{itemize}
    \item Byte Unigrams Detector. This refers to the malware detector trained using as features the byte unigrams or byte frequency distribution of the samples from the EMBER dataset.
    \item Top-K API Detector. This refers to the malware detector trained using as features the API features of the samples from the EMBER dataset.
    \item Hashed API Detector. This refers to the malware detector trained using as features the hashed version of the API features of the samples from the EMBER dataset.
    \item Hashed Strings Detector. This refers to the malware detector trained using as features the hashed version of the Strings features from the EMBER dataset.
\end{itemize}

The difference between the Top-K API and the Hashed API models is that the Top-K API models take as input a vector of 1s and 0s, indicating whether or not a particular API function has been imported. As there are millions of API functions that an executable can import, we limited the set of API functions to a subset of K functions more commonly found in benign executables, where $K \in \{150, 300, 500, 1000, 2000\}$. On the other hand, the Hashed API models use the hashing trick to vectorize the information about imported libraries and functions that can be found in the Import Address Table (IAT) of PE files into a fixed low-dimensional vector of size $1280$. Similarly, the Hashed Strings models take as input the vectorized string-based features defined in EMBER~\footnote{\url{https://github.com/elastic/ember/blob/master/ember/features.py}}. These features include statistics about the strings, their average entropy, the number of paths, urls, registries found, etcetera. For the raw string-based features, we limited the set of strings to a subset of $K$ strings more commonly found in benign executables, where $K \in \{2000, 5000, 10000\}$.

\subsubsection*{Exploring the Effects of the Gap Size on the Generated Adversarial Examples}

\begin{figure}[ht]
    \centering
    \includegraphics[width=\columnwidth]{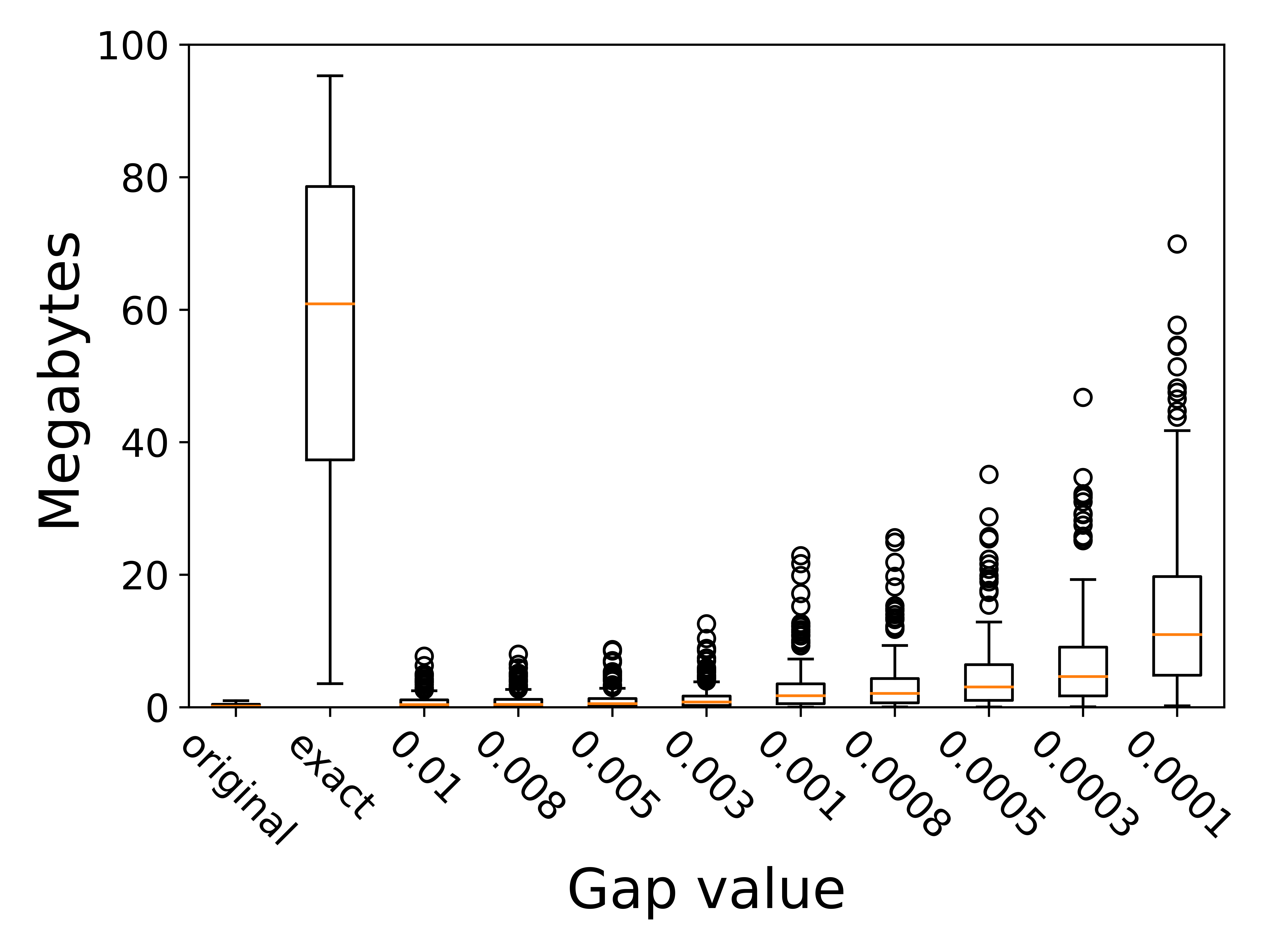}
    \caption{Size comparison of the resulting adversarial malware examples for different gap values.}
    \label{fig:boxplot_gap_size}
\end{figure}

Results in Figure~\ref{fig:boxplot_gap_size} and Table~\ref{tab:analysis_performance_gap_values} are obtained from generating adversarial malware examples on a subset of 200 samples randomly selected from the test set of the BODMAS dataset.
Appending bytes at the end of the PE executables to exactly have the target byte frequency distribution generated by the generator gives rise to large, unrealistic executables. For this reason, we have proposed to map the original byte frequency distribution to an approximated version of the target byte frequency distribution by allowing a small error. Depending on the error, the size of the adversarial examples will vary, from a few megabytes to tens of megabytes 
as shown in Figure~\ref{fig:boxplot_gap_size}.

In addition, it can be observed in Table~\ref{tab:analysis_performance_gap_values} that the greater the error between the target byte frequency distribution and the approximated version of the target byte frequency distribution the greater the accuracy of the byte-based unimodal detector on the resulting adversarial examples. This is because the greater the error the greater the differences between the generated byte frequency distribution and its approximated version. Table~\ref{tab:analysis_performance_gap_values} shows that the accuracy of the models trained on the byte unigram features drops to 0\% for the exact solutions. However, the resulting malicious executables are non-viable as they have an average size of $\approx$56MB, a 20918.5\% increase with respect to the original executables. A good trade-off between evasion rate and the size of the adversarial examples is observed using 0.001 as the gap value or error. A higher gap generates less evasive adversarial examples while a lower gap generates adversarial examples that are too big compared to the original size of the executables. Thus, for the remaining experiments, the approximated target byte frequency distributions will be generated using a gap value equal to 0.001.

\begin{table*}[tb]
\centering
\caption{Detection rate of the Byte-based malware detector against the adversarial malware examples generated by the GAN using various gap values. The size of the executables in Megabytes is the average over the set.}
\label{tab:analysis_performance_gap_values}
\resizebox{\textwidth}{!}
{%
\begin{tabular}{l|cc|ccccccccc}
\hline
\multirow{2}{*}{Detector} &       &       & \multicolumn{9}{c}{Gap}                                                  \\ \cline{2-12} 
                          & Orig. & Exact & 0.01 & 0.008 & 0.005 & 0.003 & 0.001 & 0.0008 & 0.0005 & 0.0003 & 0.0001 \\ \hline
Byte Unigram                & 0.895  & 0.0    & 0.48  & 0.385  & 0.3    & 0.1    & 0.005  & 0.015 & 0.0     & 0.0     & 0.0     \\ \hline
MB              & 0.27  & 56.75 & 0.89 & 0.96  & 1.15  & 1.47  & 2.99  & 3.53   & 4.99   & 7.19   & 14.45  \\ \hline
\end{tabular}
}
\end{table*}    

\subsubsection*{Comparison with MalGAN and the Benign Code Injection Attack}

This section presents a comparison of our query-free attacks against the benign code injection attack and MalGAN~\cite{DBLP:journals/corr/HuT17}.

On the one hand, the Benign Code Injection Attack is a well-known attack against ML-based malware detectors that consists of injecting benign content within the malicious content to try to disguise the malicious code and make it look more like benign code. This attack serves as a baseline to evaluate the feasibility of our query-free attack based on GANs as it is the only attack presented in the literature that can be implemented without querying the target malware detectors. To this end,
we generate adversarial malware examples by injecting the code of a randomly selected benign example into its overlay. Different variations of the benign code injection attack exist, i.e. create one or more new sections with the benign content, etcetera, but for simplicity purposes we decided to just append the benign content at the end of the file.

On the other hand, MalGan is state-of-the-art GAN-based approach to generate adversarial malware examples. MalGAN consists of two feed-forward neural networks, (1) a generator and (2) a substitute detector. The generator
network is trained to minimize the generated adversarial malware examples’ maliciousness probabilities predicted by the substitute detector whereas the substitute detector is trained to fit the black-box malware detection system. MalGAN was originally trained on a 160-dimensional binary feature vector for each program, based on 160 system level APIs, and evaluated on various black-box detectors, i.e. random forest, logistic regression, decision trees, etc, trained on the same feature set. However, MalGAN relies on having unrestricted access to the black-box detection system to be able to train a good substitute detector. In contrast, our attack does not require querying the black-box detection system at all. Notice that MalGAN was proposed for evading API-based detectors. In this work, we adapted and extended MalGAN to also attack byte-based and string-based malware detectors.

\begin{figure}[h]
        \centering
        \includegraphics[width=\columnwidth]{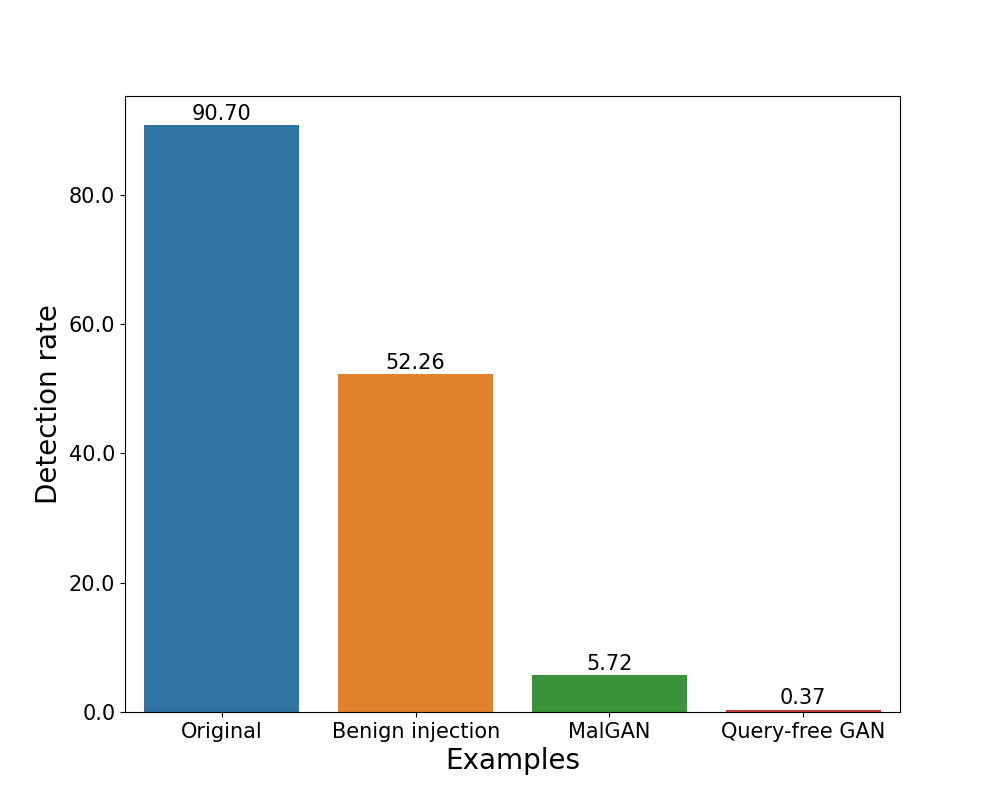}
        \caption{Detection rate of the byte-based detectors on the original and the adversarial byte histogram features.}
        \label{fig:evasion_against_byte_detectors}
\end{figure}

Figure~\ref{fig:evasion_against_byte_detectors} presents the detection rate of the byte-based ML model on the adversarial examples generated by the benign code injection attack, MalGAN and our approach. It can be observed that all approaches reduced the detection rate of the target classifier from approximately 90.70\% to 52.46\%, 5.72\% and 0.37\%, respectively. 
It is important to note that among all three methods, the adversarial samples produced by the benign code injection attack are the least evasive. This is due to the fact that even though adding benign content changes the malicious executable's byte frequency distribution, doing so requires adding a significant amount of content compared to the executable's original size in order to flip the classifier's prediction from a "malicious" to a "benign" byte frequency distribution.
Notice that both MalGAN and our query-free approach successfully reduce the detection rate of the ML-based detector to almost zero. On the one hand, by non-restrictively interacting with the target malware classifier, MalGAN learns to which target byte frequency distribution the executables must be mapped in order to flip the classifier's prediction. On the other hand, our query-free approach discovers which byte frequency distribution corresponds to "genuine" or "benign" executables. Results suggest that if the features can be freely modified without restrictions, altering the features in a way that they look "benign" is a plausible way to evade detection. However, having access to the same feature set as the ML detection model is a best case scenario. 

\begin{figure*}[ht]
    \centering
    \begin{subfigure}[h]{\columnwidth}
        \centering
        \includegraphics[width=\columnwidth]{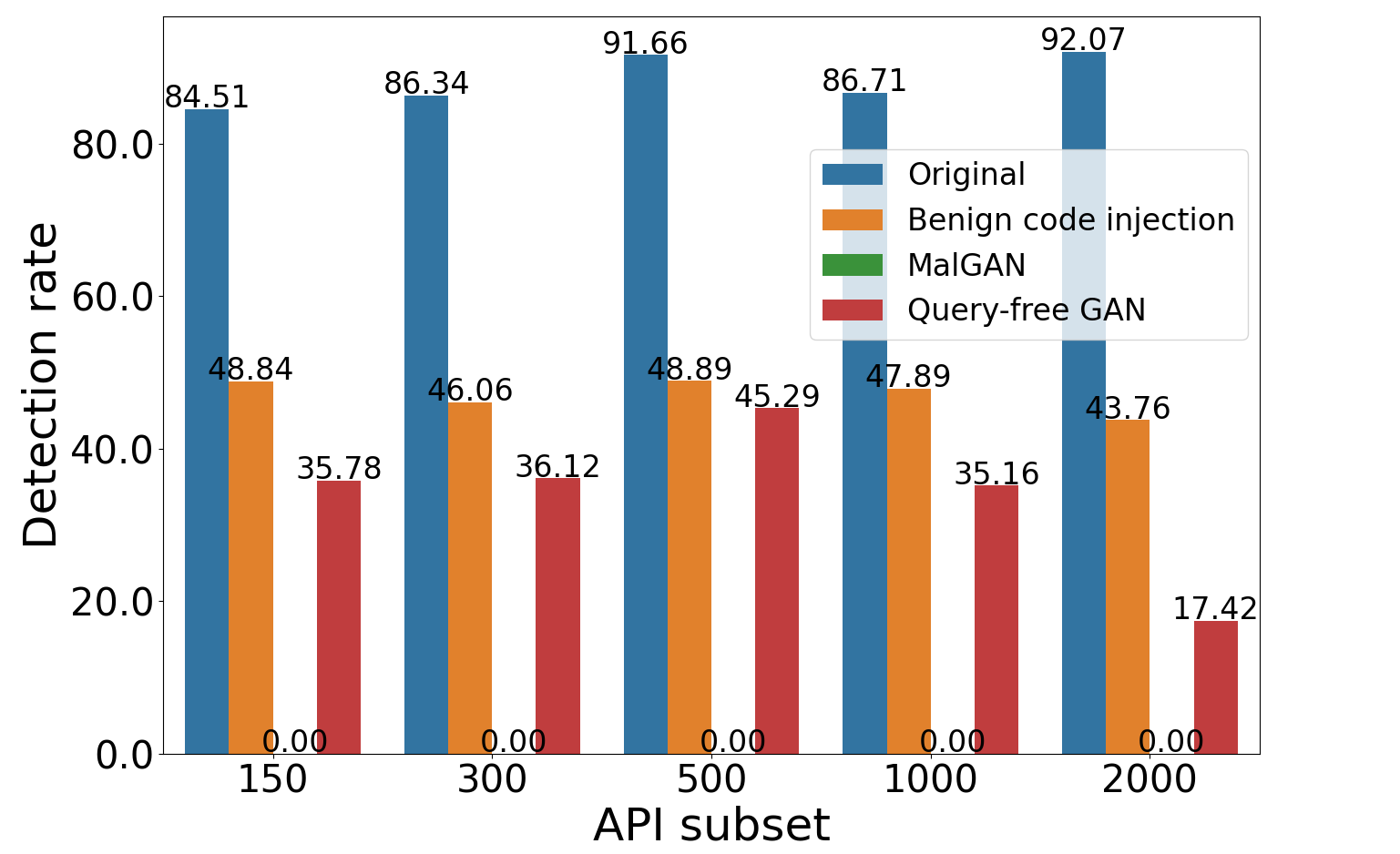}
        \caption{Detection rate of the Top-K API-based detectors on the adversarial API-based feature vectors}
        \label{fig:detection_rate_top_k_api}
    \end{subfigure}
    \hfill
    \begin{subfigure}[h]{\columnwidth}
        \centering
        \includegraphics[width=\columnwidth]{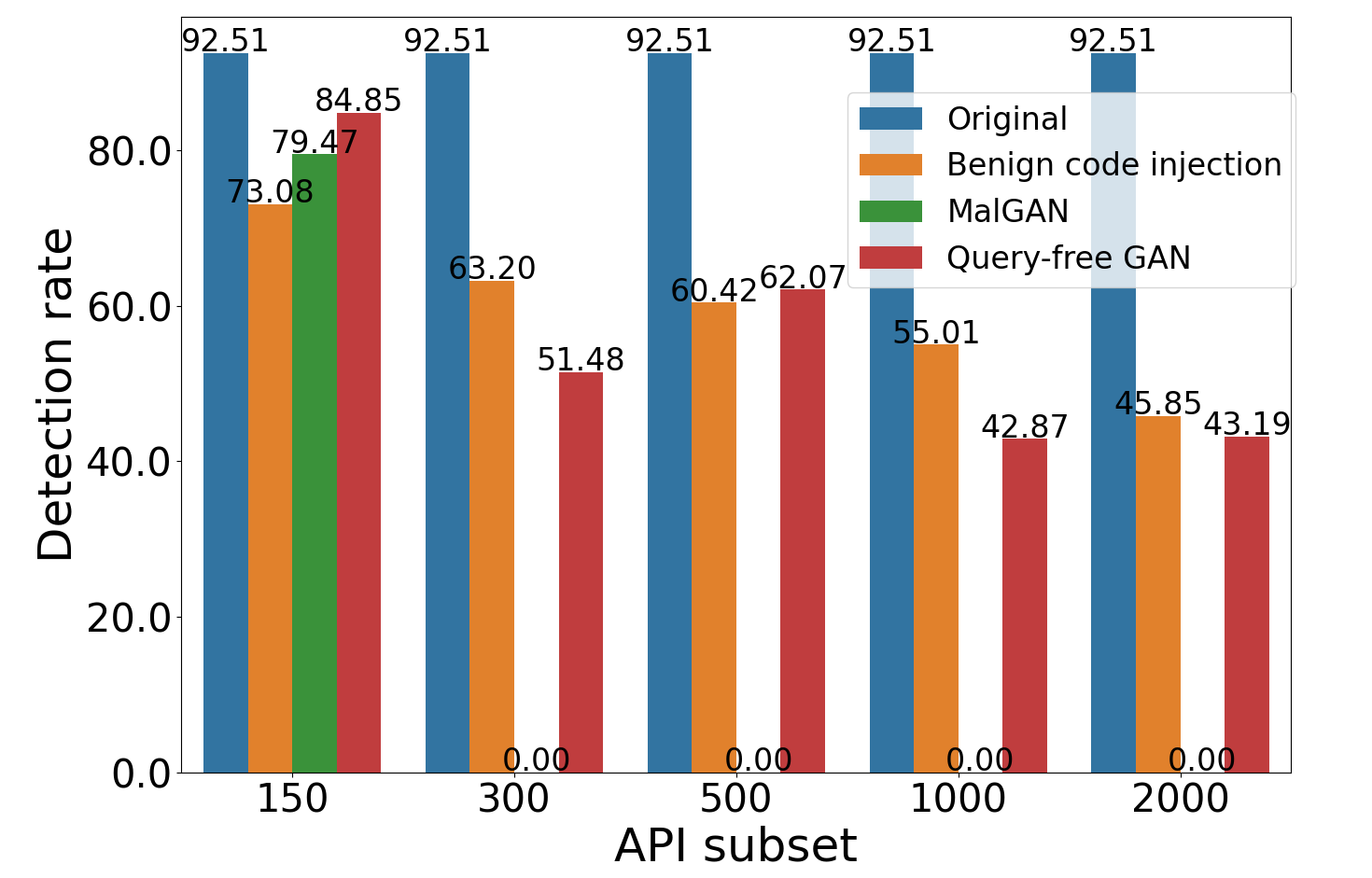}
        \caption{Detection rate of the Top-K Hashed API-based detectors on the adversarial API-based feature vectors}
        \label{fig:detection_rate_hashed_top_k_api}
    \end{subfigure}
    \hfill
    \begin{subfigure}[h]{\columnwidth}
        \centering
        \includegraphics[width=\columnwidth]{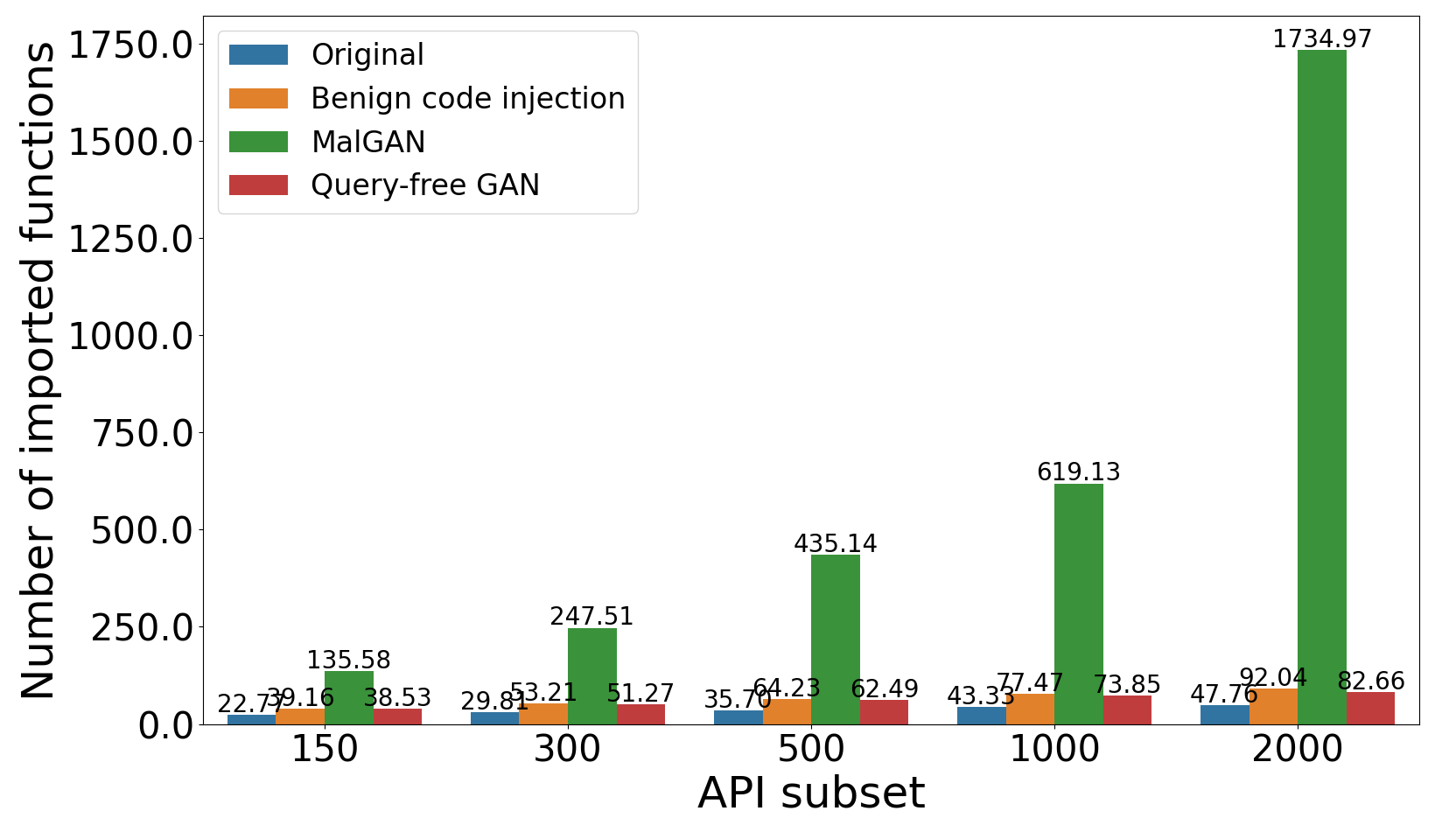}
        \caption{Number of functions imported by the adversarial API-based feature vectors.}
        \label{fig:num_function_top_k_api}
    \end{subfigure}
    \caption{Detection rate of the Top-K and Hashed API-based detectors on the original and the adversarial API-based features.}
    \label{fig:detection_rate_top_k_api_gan}
\end{figure*}


\begin{figure*}[ht]
    \centering
    \begin{subfigure}[h]{\columnwidth}
        \centering
        \includegraphics[width=\columnwidth]{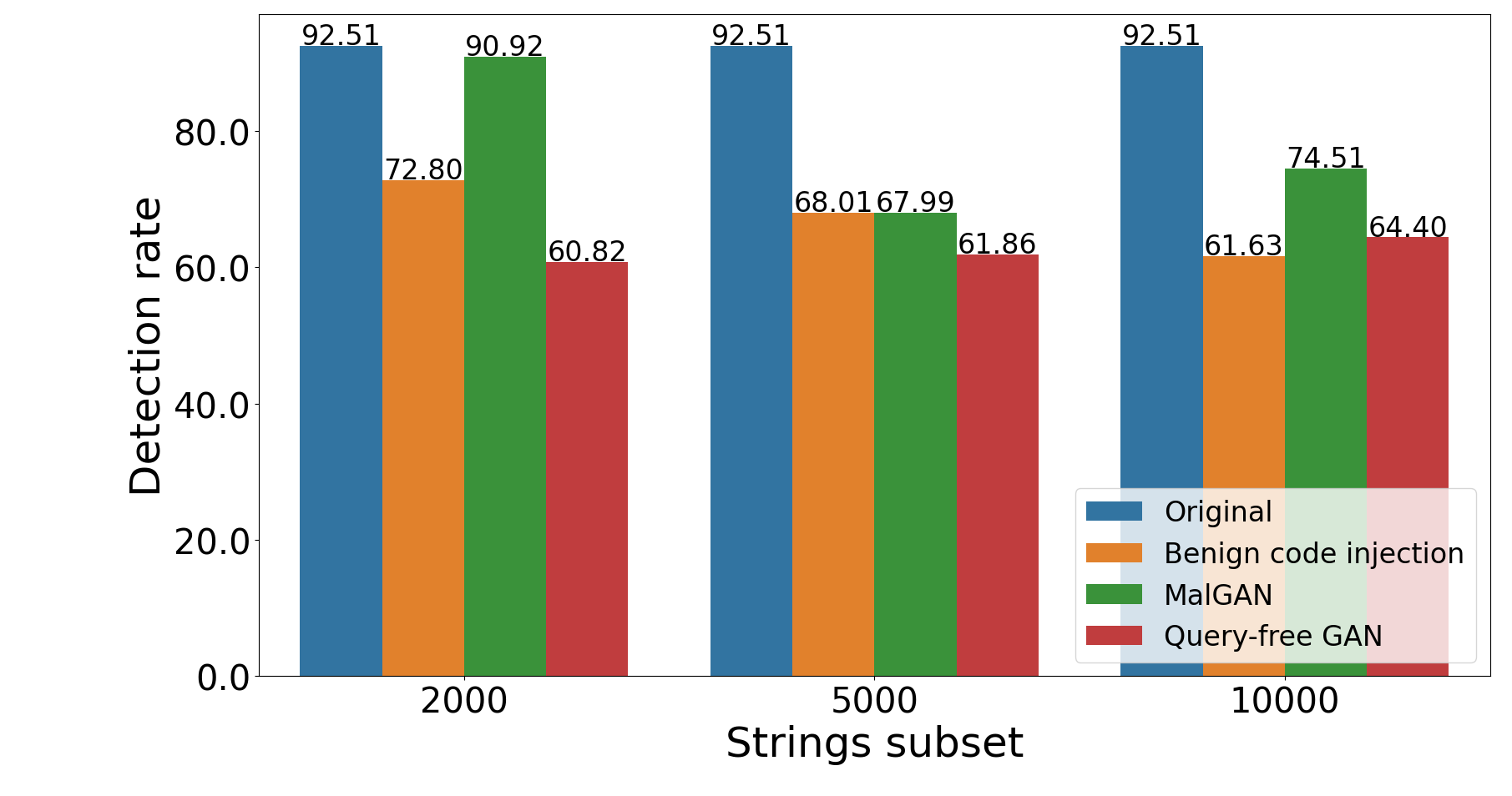}
        \caption{Detection rate of the Top-K String-based detectors on the adversarial String-based feature vectors}
        \label{fig:detection_rate_top_k_strings}
    \end{subfigure}
    \hfill
    \begin{subfigure}[h]{\columnwidth}
        \centering
        \includegraphics[width=\columnwidth]{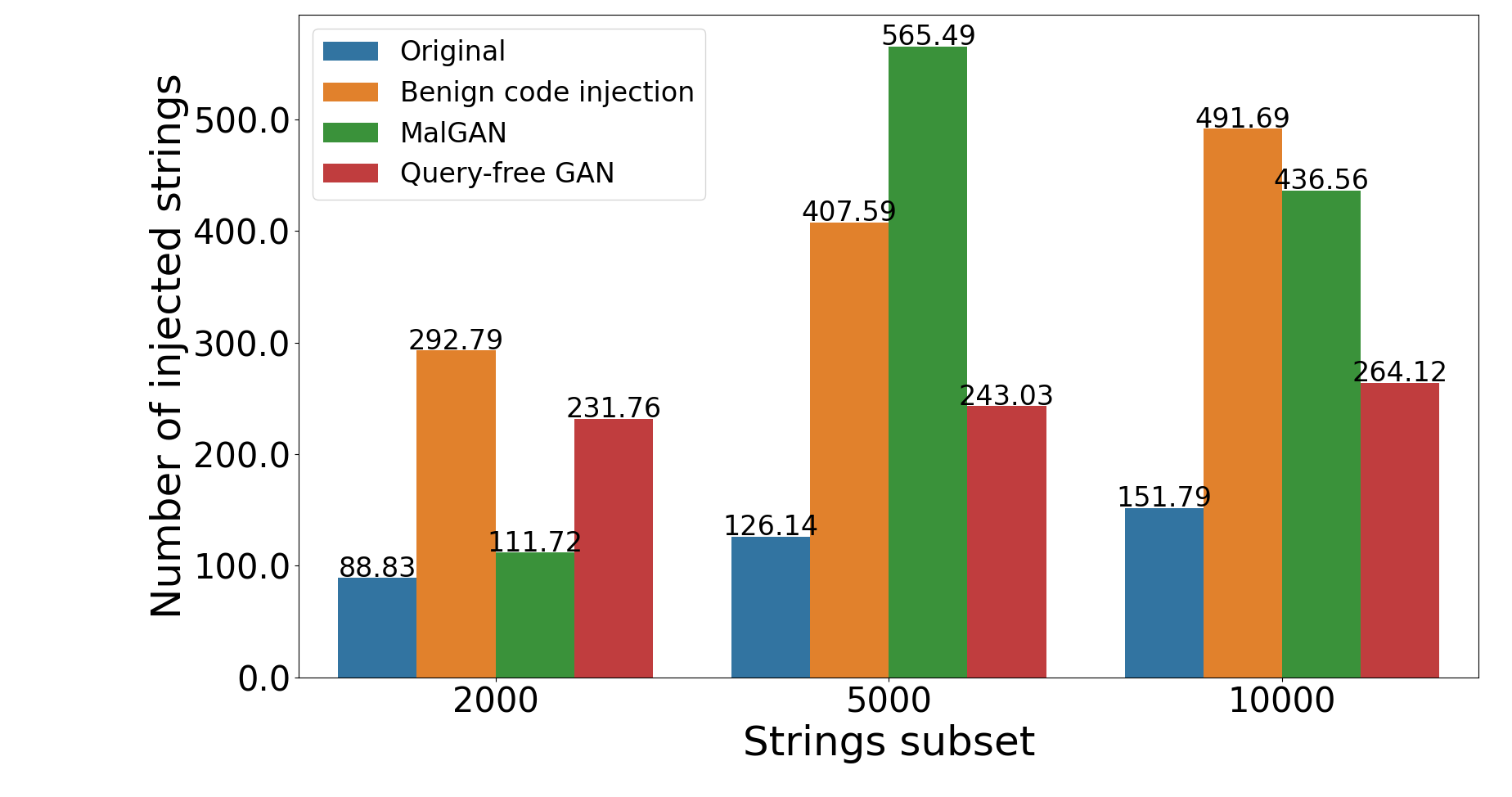}
        \caption{Number of strings injected by the adversarial String-based feature vectors.}
        \label{fig:num_function_top_k_strings}
    \end{subfigure}
    \caption{Detection rate of the Top-K Hashed String-based detectors on the original and the adversarial String-based features.}
    \label{fig:detection_rate_top_k_strings_gan}
\end{figure*}
A more common scenario is to only have access to the raw features used by the ML model instead of the vectorized features or the final representation of those features. For instance, let us consider the list of API libraries and functions imported by Portable Executable files. This information can be obtained from the Import Address Table. In this case, we have access to the raw features, or the list of libraries and functions imported, but we may not be aware of the exact process that has been used to produce the final feature vector representation.
 As there are millions of API libraries and functions it is common to employ feature selection and dimensionality reduction techniques to map the high-dimensional vectors to a low-dimensional representation that is later used to train the ML models. In the present case, we are still able to modify the raw features by importing new libraries and functions but the effect of such modifications may be constrained in some way by the feature selection and dimensionality reduction techniques employed to map the high-dimensional API-based features to a more tractable low-dimensional representation, even though some transferability will apply. Furthermore, not all features can be adjusted without restraint. For instance, existing API features, i.e. the presence of a particular API function in the program, cannot be removed as that would make the malware crack. Thus, only new features can be added.

Figure~\ref{fig:detection_rate_top_k_api_gan} presents the detection rate of various ML models trained on the raw API-based features and their hashed representation, respectively. On the one hand, the raw API-based feature vector is a vector $x$ of size $M$, where each element in $x$ indicates whether or not a particular API function has been imported. On the other hand, the hashed API feature vector $x^{'}$ maps the information about the imported libraries and functions from the Import Address Table to a low-dimensional feature vector using the hashing trick. For detailed information on the mapping between the raw and hashed features we refer the reader to the work of H. Anderson et al.~\cite{2018arXiv180404637A}. 

In Figure~\ref{fig:detection_rate_top_k_api_gan}, it can be observed that the percentage of detected adversarial examples generated by our approach is greater than that of the MalGAN approach. Our intuition is that because "malicious" features cannot be removed and only new features can be added it is more difficult to disguise the feature vectors as "benign". Another complementary explanation is that there is a great deal of overlapping between the API libraries and functions used by benign and malicious software and thus, importing "benign" functions is not the best choice as those functions are also found in malicious software. It should be noted that MalGAN achieves 100\% evasion by generating adversarial examples that could be considered outliers as they are importing an unrealistic amount of libraries and functions from the Windows Application Programming Interface. In contrast to the 47.76 functions imported in average by the original malicious executables, MalGAN's adversarial feature vectors import an average of 1751.91 and 1734.97 functions. This represents a 3568.15\% and 3532.68\% increase, respectively. 
In comparison, our approach generates more "real" feature vectors while importing only 82.66 functions on average per sample, less than twice the number of functions imported in the original samples. Furthermore, it can be observed that the detection rate of the ML models trained on the hashed API features is higher than the one reported by the model trained on the raw API features suggesting that even though the modifications performed on the raw features are transferred to the hashed features, the hashed representation reduces the effectivity of the alterations. 

Lastly, we present the detection rate of the hashed string-based ML models in Figure~\ref{fig:detection_rate_top_k_strings_gan}. In this case, neither approach is able to reduce the detection rate of the String-based model below 60\%. Notice that MalGAN is not able successfully generate adversarial examples similarly to when the target detector was trained on the hashed API-based features. The reason behind is that the string-based vectorized feature vector contains statistics about strings, their entropy, number of paths, urls, etcetera, and thus, it is difficult for MalGAN to learn how injecting strings affects the prediction of the target classifier. In contrast, our query-free approach injects strings so that the raw feature vector representation looks "benign" rather than relying on the feedback of the target classifier. However, injecting strings is not enough to evade the String-based model as it not only uses features related to the ASCII strings found in the executable's content but the number of paths, urls, registries, number of printables, etcetera, which are not altered by injecting strings. 

Notice that the evasion rates of the proposed query-free approach are greater when we allow the generator to select among the 2000 most used APIs and Strings found in benign samples. For this reason, in the next Section we use the corresponding generators to generate the adversarial examples.



\subsubsection{Attack Evaluation}

\subsubsection{Attack Evaluation on State-Of-The-Art Malware Detectors}
Next, the quality of the adversarial examples generated by our query-free model-agnostic GAN, will be assessed against the following state-of-the-art detectors: (1) the EMBER LightGBM model and (2) the MalConv model.

The EMBER model refers a a gradient boosting trees model (feature-based detector) that receives as input a feature vector consisting of the following types of features:
\begin{itemize}
    \item Byte unigram features.
    \item 2D byte/entropy histogram features~\cite{DBLP:conf/malware/SaxeB15}.
    \item Information about the section names, their sizes and entropy.
    \item Information about the imported libraries and functions from the Import Address Table (IAT).
    \item Information about the exported functions.
    \item General information about the file such as its size, the virtual size, the number of imported and exported functions, whether it has a signature, etcetera.
    \item Information extracted from the header such as the targeted machine, its architecture, OS, the major and minor linker versions, etcetera.
    \item Information about the strings extracted from the raw byte stream.
    \item Information about the size and virtual address of the first 15 data directories.
\end{itemize}
The resulting feature vector has size 2351, where 256, 1280, and 104 of the features correspond to the byte unigram features, the API features, and the string features, respectively. In contrast, the MalConv model refers to a shallow convolutional neural network (deep learning-based detector) which receives as input the raw byte stream, up to 1,048,576 $\sim$1Mb. 

In addition to the aforementioned state-of-the-art detectors we trained two multimodal detectors containing (1) the byte and API-based features and (2) the byte, API, and String-based features. These models are referred as EMBER v1 and EMBER v2 from now on. Notice that all models have been trained using the data from the EMBER dataset~\cite{2018arXiv180404637A}.

\begin{table*}[ht]
\caption{Detection rate of SOTA ML-based detectors on the adversarial examples.}
\label{tab:detection_rate_sota_ml_detectors}
\resizebox{\textwidth}{!}{%
\begin{tabular}{l|l|ccccl}
\hline
\multirow{2}{*}{ML Detector} & \multirow{2}{*}{Original examples} & \multicolumn{5}{c}{Adversarial examples}                                                                                                                                              \\ \cline{3-7} 
                             &                                    & \multicolumn{1}{c|}{Byte-based GAN} & \multicolumn{1}{c|}{API-based GAN} & \multicolumn{1}{c|}{String-based GAN} & \multicolumn{1}{c|}{Byte+API+Strings-based GAN} & Benign Injection \\ \hline
EMBER v1                     & 97.02                              & \multicolumn{1}{c|}{32.93}          & \multicolumn{1}{c|}{87.23}         & \multicolumn{1}{c|}{90.67}            & \multicolumn{1}{c|}{\textbf{28.48}}                      & 52.81            \\
EMBER v2                     & 96.03                              & \multicolumn{1}{c|}{9.36}           & \multicolumn{1}{c|}{88.17}         & \multicolumn{1}{c|}{85.20}            & \multicolumn{1}{c|}{\textbf{8.15}}                       & 51.22          \\
EMBER                        & 98.64                              & \multicolumn{1}{c|}{83.16}          & \multicolumn{1}{c|}{99.11}         & \multicolumn{1}{c|}{96.45}            & \multicolumn{1}{c|}{\textbf{82.84}}                      & 96.55            \\
MalConv                      & 91.34                              & \multicolumn{1}{c|}{\textbf{53.74}}          & \multicolumn{1}{c|}{89.14}         & \multicolumn{1}{c|}{86.61}            & \multicolumn{1}{c|}{72.10}                      & 70.13             \\ \hline
\end{tabular}%
}
\end{table*}

The proposed attack is model-agnostic, i.e. does not require knowing anything about the target malware detectors. However,
the attack requires that the target malware detectors are influenced directly or indirectly by the features modified. For instance, by appending bytes at the end of the executable in order to move from the original to the target byte frequency distribution we will indirectly modify other features from the EMBER set such as the 2D byte/entropy histogram features, the size of the file, and the size of the overlay. Similarly, if we modify the Import Address Table to inject new API libraries and functions, the size of the file, the number of imported functions and other features from the EMBER set will be indirectly modified. The same occurs when adding a new section containing the benign strings we want to inject. In this case, the Section Table will be modified with a new entry, new content will be added to the newly-created section, and thus, some features will be indirectly modified in addition to the strings features of the EMBER set. Note that the aforementioned modifications to the executables manipulate their byte contents and thus, the feature-based attacks might also transfer to deep learning models that take as input the raw byte sequence of executables, i.e. MalConv.

The detection accuracy of the ML-based detectors on the test set's original and generated adversarial examples is shown in Table~\ref{tab:detection_rate_sota_ml_detectors}. It can be observed that the adversarial examples generated by the byte-based GAN decrease the detection accuracy of the deep learning model, i.e. MalConv, from 91.34\% to 53.74\%, respectively.  We conclude that this interesting drop in performance is because MalConv learned that large chunks of a given byte value, such as the perturbations we perform, are indicative of benign samples. In addition, the results show that when combining the three types of modifications the detection accuracy of the generated adversarial examples drops even more, showing that the more modifications that are stacked together the lower the detection accuracy, as more features will be modified. This applies to any feature-based detector, i.e. EMBER, EMBER v1, and EMBER v2. Furthermore, the findings in Table~\ref{tab:detection_rate_sota_ml_detectors} indicate that our query-free approach generates more evasive adversarial examples than the benign code injection attack, decreasing the detection accuracy of the EMBER v1, EMBER v2 and EMBER models from 97.02\%, 96.03\% and 98.64\% to 28.48\%, 8.15\%, and 82.84\%. In contrast, the benign code injection attack is unsuccessful in reducing the performance of the EMBER models down to less than 50\%. We would like to point out that we cannot provide a comparison of MalGAN and our approach against the
SOTA ML detectors in Table~\ref{tab:detection_rate_sota_ml_detectors}. The reason is that MalGAN generates very anomalous API-based feature vectors and when we tried to map those feature vectors back to the executables the server ran out of resources.

\subsubsection{Attack Evaluation on VirusTotal Service}
The adversarial malware executables generated by our attack were uploaded to VirusTotal to check whether or not the number of detections decreased in comparison to the detections in the original executables. We would like to point out that the adversarial examples have been specifically generated to evade a feature-based ML detectors by modifying the feature vectors so as they look "benign" and thus, it is unrealistic to expect the adversarial examples to evade real-world malware detectors. Nevertheless, results show that the generated adversarial examples are able to evade various anti-malware engines.

Table~\ref{tab:vt_detections} presents the median number of average VirusTotal detections in a subset of samples randomly selected from the test set. It an be observed that the average number of detections decreases from $59.05/74$ to $52.95/74$, $47.4/74$ and $50.05/74$, for the adversarial examples generated by appending bytes at the overlay, importing new API functions into the Import Address Table and injecting strings into a new section, respectively. Furthermore, by stacking various types of modifications the average number of detections in the adversarial examples decreases to $46.57/74$. Results suggest that the more modifications applied to the original samples the higher the evasion rate. 

\begin{table}[ht]
    \centering
    \caption{Cross-evasion rates on 200 randomly chosen holdout samples from the test set, showing the median number of VirusTotal detections of the original executables and the adversarial executables.}
    \label{tab:vt_detections}
    \begin{tabular}{l|c}
    \hline
    Type of Examples & Number of Detections \\ \hline
    Original        & 59.05/74             \\
    Byte histogram  & 52.95/74             \\
    API imports     & 47.4/74              \\
    Strings         & 50.05/74             \\
    Combination     & 46.57/74             \\ \hline
    \end{tabular}
\end{table}

\section{Conclusions}
\label{sec:conclusions}
Recent research on evasion attacks against ML-based malware detectors is limited and impractical in a real-world scenario where no knowledge about the detection system is available and the attackers do not have unlimited queries to the detection system. This paper presents the first model-agnostic attack that generates adversarial malware examples without querying the detection system and without assuming partial or complete knowledge of the system. The proposed attack modifies the malicious executables in a way that make them look benign, and thus makes them harder to detect by malware detection systems. This represents a novel and unexplored direction in automatic evasion research.

\subsection{Discussions}
\label{sec:discussions}
ML-based malware detectors have been proven to be susceptible to evasion attacks. However, existing white-box and black-box attacks require access to some sort of information about the ML detector in order to succeed, i.e. the algorithm used to train the ML detector, its parameters, its output, etcetera. This limits the applicability of the aforementioned attacks in the real-world as information about the ML models or the scores associated with a submitted executable might not be available. To circumvent the limitations of existing approaches, we designed a GAN-based framework that generates adversarial examples by modifying the malware's features so they resemble those found in benign executables. 

This work provides an alternative approach to generate adversarial malware examples when restrictions are imposed on access to the model's algorithm, parameters and number of queries. In the hypothetical scenario where malware authors have access to the model's training algorithm, parameters, scores and have unlimited queries, any existing evasion attack might perform better than our approach. In general, independently of the domain application, the evasion rate of the adversarial examples generated with white-box attacks is greater than that of those generated by black-box attacks as they can use the model's information and gradients to tweak the adversarial examples at their convenience. Furthermore, adversarial examples generated with score-based attacks are usually more evasive than than label-based attacks as the scores can be used to numerically estimate the gradient. Lastly, label-based attacks will generate more deceptive examples than those that do not use any kind of output from the model to generate the adversarial examples. This is true for all domain applications and not only for the task of malware detection. Thus, it is unrealistic to expect a query-free attack to generate more elusive adversarial examples than those evasion attacks that use any kind of information and output from the ML detector.

Nevertheless, experiments have shown that the adversarial examples generated by our approach achieve similar evasion rates to MalGAN but without needing to query the target ML detector in the process. In addition, the adversarial examples generated by our approach look more \emph{real} than the ones generated by MalGAN, as the later could well be labelled as outliers. In addition, results show that the query-free approach decreases the detection accuracy in unimodal detectors as well as the accuracy of multimodal and deep learning detectors. Moreover, results suggest that stacking one or more modifications together leads to better evasion rates. 

Furthermore, we have made every effort to evaluate the generated adversarial examples against the widest range of state-of-the-art malware detectors as possible, including multimodal and deep learning detectors. 
We believe our approach to generate adversarial examples starts a new methodology, and that it is more challenging than the approaches experimentally compared with in this work. 
For helping further research, we have open-sourced our code under the MIT License and we have used a public benchmark to evaluate our approach in order to allow researchers to reproduce our work and build upon it.

\subsection{Future Work}
\label{sec:future_work}
The proposed query-free attack has been applied to modify three types of features commonly employed to detect malware: (1) byte unigram features, (2) API-based, and (3) string-based features. Apart from the aforementioned features, static ML-based malware detectors employ a wide range of feature types, i.e. information extracted from the PE header, the list of exported functions, the properties of each section, byte and opcode n-gram features, etc. Thus, a natural extension of our work would be to extend our approach to deal with these features. In addition, our approach could be used in conjunction with the typically employed compression and encryption techniques used to obfuscate the malware examples, and in conjunction with the newer adversarial evasion techniques developed to bypass ML-based detection.

\section*{Acknowledgements}
This project has received funding from the Spanish Science and Innovation Ministry funded project PID2019-111544GB-C22, Enterprise Ireland and the European Union’s Horizon 2020 Research and Innovation Programme under Marie Skłodowska-Curie grant agreement No 847402. The views and conclusions contained in this document are those of the authors and should not be interpreted as representing the official policies, either expressed or implied, of CeADAR, University College Dublin, IBM Ireland Limited, and the University of Lleida. We would like to thank Cormac Doherty and UCD's Centre for Cybersecurity and Cybercrime Investigation for their support.

\section*{Data and Code Availability}
The BODMAS dataset is available to the public~\footnote{\url{https://whyisyoung.github.io/BODMAS/}} and the source code~\footnote{\url{https://github.com/code_repository}}
of our approach will be made available under a MIT License after the paper is accepted.

\bibliographystyle{plain}
\bibliography{references}

\appendices
\section{Configuration Details of the Generator and Critic Networks}

A grid search was used to select the appropriate hyperparameters of the generator and discriminator networks. The following configuration details refer to the best byte-based architectures. The difference between the byte-based architectures and the API-based and string-based architectures is two-fold: (1) the number of the input features, i.e. 256, 2000 and 2000, respectively; and (2) the number of hidden neurons in the generator and critic architectures, [256, 256, 1] and [128, 64, 1], [2000, 2000, 1] and [500, 300, 100, 1], and [512, 512, 1] and [500, 300, 100, 1], for the byte-based, API-based and string-based generator and critic architectures, respectively. Notice that every network architecture starts with a dropout layer. The reason behind starting with a dropout layer following the input layer is to prevent overfitting. The BODMAS dataset only contains 134435 executables and thus, the models trained on it are prone to overfitting. Dropping out a subset of the input features as well as dropping out neurons from the hidden layers helps avoiding overfitting. The first dropout layer drops out 10\% of the neurons while the hidden dropout layers drop out 50\% of the neurons. 

\subsection{Byte Unigram Features Generator and Critic Networks}
\begin{table}[h]
	\centering
	\caption{Configuration details of the generator network. M is the size of the byte unigram feature vector, $M=256$. Z is the size of the random noise vector, $Z=8$ }
	\label{tab:unigram_generator_configurator_details}
	\begin{tabular}{lll}
		\hline
		Layer (type)                           & Output shape   & Parameters \# \\ \hline
		input (Input layer)                    & (batch\_size, M+Z)   & 0             \\
		dropout\_1 (Dropout layer)             & (batch\_size, M+Z)   & 0             \\
        dense\_1 (Dense layer)                 & (batch\_size, 256)   & (M+Z)*256 \\
        relu\_1 (ReLU layer)                 & (batch\_size, 256)   & 0\\
        dropout\_2 (Dropout layer)             & (batch\_size, 256)   & 0             \\
        dense\_2 (Dense layer)                 & (batch\_size, 256)   & 256*256 \\
        relu\_2 (ReLU layer)                 & (batch\_size, 256)   & 0\\
        dropout\_3 (Dropout layer)             & (batch\_size, 256)   & 0             \\
        dense\_3 (Dense layer)                 & (batch\_size, 1)     & 256*1             \\
        softmax\_1 (Softmax)                   & (None, N)            & 0             \\ \hline
		\multicolumn{2}{l}{Total trainable parameters}                & 133376             \\ \hline
	\end{tabular}%
\end{table}

\begin{table}[h]
	\centering
	\caption{Configuration details of the critic network. M is the size of the byte unigram feature vector, $M=256$.}
	\label{tab:unigram_critic_configurator_details}
	\begin{tabular}{lll}
		\hline
		Layer (type)                           & Output shape   & Parameters \# \\ \hline
		input (Input layer)                    & (batch\_size, M)   & 0             \\
		dropout\_1 (Dropout layer)             & (batch\_size, M)   & 0             \\
        dense\_1 (Dense layer)                 & (batch\_size, 128) & M*128 \\
        leaky\_relu\_1 (Leaky ReLU layer)                   & (batch\_size, 128)   & 0\\
        dropout\_2 (Dropout layer)             & (batch\_size, 128) & 0             \\
        dense\_2 (Dense layer)                 & (batch\_size, 64)  & 128*64 \\
        leaky\_relu\_2 (Leaky ReLU layer)                 & (batch\_size, 64)   & 0\\
        dropout\_3 (Dropout layer)             & (batch\_size, 64)  & 0             \\
        dense\_3 (Dense layer)                 & (batch\_size, 1)   & 64*1             \\
		\multicolumn{2}{l}{Total trainable parameters}              & 41024             \\ \hline
	\end{tabular}%
\end{table}

\end{document}